\documentclass{article}
\usepackage{amsmath}
\usepackage{graphicx}
\usepackage{amsfonts}
\usepackage{amssymb}
\usepackage{geometry}
\usepackage{float}
\usepackage{bm}
\usepackage{color}
\usepackage{cite}
\usepackage{subfigure}
\usepackage{geometry}
\usepackage{geometry}

\begin{document}

\title{The effect of modified dispersion relations on the thermodynamics of
Schwarzschild black hole surrounded by quintessence}
\author{B. Hamil\thanks{%
hamilbilel@gmail.com } \\
%EndAName
D\'{e}partement de TC de SNV, Universit\'{e} Hassiba Benbouali, Chlef,
Algeria. \and B. C. L\"{u}tf\"{u}o\u{g}lu\thanks{%
bekircanlutfuoglu@gmail.com (corresponding author)} \\
%EndAName
Department of Physics, University of Hradec Kr\'{a}lov\'{e}, \\
Rokitansk\'{e}ho 62, 500 03 Hradec Kr\'{a}lov\'{e}, Czechia.}
\date{}
\maketitle

\begin{abstract}
In this manuscript, we investigate the effects of a modified dispersion relation on the thermodynamics of Schwarzschild black hole surrounded by the quintessence matter. We find that the MDR-correction states the same lower bound on the horizon, while the quintessence matter specifies the upper bound to the horizon depending on the state parameter. Due to MDR-correction and quintessence matter presence, we observe modifications in equation of state and specific heat functions of black hole. We show that a remnant can occur according to quintessence matter, and the black hole's stability depends only on the modified dispersion relation. 
\end{abstract}

\section{Introduction}
Planck mass, $M_p$, and Planck length, $\ell_p$, concepts, which are universal invariants, are indispensable for quantum gravity theories. According to these theories,  in the Planck scale space time is not continuous, and thus, particles can not be localized smaller than the Planck length \cite{Gross, Maggiore}. However then, one of the cornerstone principles of general relativity, the 
Lorentz symmetry, can not be preserved \cite{cam01}. Considering a deformation in the ordinary dispersion relation, namely the modified dispersion relation (MDR), can obscure this important flaw, and restore doubly special relativity (DSR) \cite{Magueijo1, Magueijo2}. Black holes physics, like String Theory and DSR, predicts the existence of a measurable minimum length \cite{ali21}. Especially, in the Planck scale quantum gravity effects  gain importance in black hole physics \cite{Gaddam}.  In many papers it is shown that the MDR has a great impact on the thermal quantities of black holes \cite{Cam03, Camelia}, particularly in black hole evaporation \cite{ Ling, Han, Kamali01, Kamali02, Lobo, Sepangi}. 

From observational astronomical evidence, we know that the universe is expanding at an accelerating rate \cite{Riess1,Riess2,Perlmutter1999}. We can use the dark energy concept as justification for the acceleration. According to our current understanding, dark energy is a negative pressure formed of energy and is spread in the whole universe by occupying around $70\%$ of the total energy density. Defining a cosmological constant is the easiest mathematical way to model the dark energy \cite{Padman2003}. Unfortunately, its experimental value was observed quite smaller than its expected value theoretically \cite{SW1989}. For this reason physicists proposed alternative models based on dynamic scalar fields differing with a parameter value showing the ratio of pressure {\color{red}of} dark energy to energy density \cite{carroll1998, Khoury2004, Picon2000, Padman2002, Caldwell2002, Gasperini2002}. One can achieve a deeper analyze and comparison of the models in the review paper of Copeland et al.  \cite{Copeland2006}.

Two decades ago,  Kiselev employed the quintessence matter model, as one of the {\color{red}dark} energy candidate, to investigate the thermodynamics of the Schwarzschild black hole \cite{Kiselev}. Later, Chen et al. examined the Hawking radiation in a $d$-dimensional spherically symmetric static  black hole surrounded by  quintessence matter \cite{Chen2008}. Then, Wei et al. presented the impact of quintessence matter on the thermal quantities for the Reissner- Nördstrom black holes \cite{Wei2011, Wei2013}. Meanwhile, Narai type black holes surrounded by quintessence matter are investigated by Fernando \cite{Fernanado2013a, Fernanado2013b}. Soon after Ghaderi et al. explored the thermodynamics functions 
of quintessence matter surrounded Bardeen and Schwarzschild black holes  \cite{Ghaderi2016a, Ghaderi2016b}. Shahjalal contemplated quantum corrected Schwarzschild metric and investigated the thermal quantities of the black hole with and without the quintessence matter \cite{3}. {\color{red} Nozari et al. examined the tunnelling process for massive and massless  particles of quantum deformed Schwarzschild  black hole in the background of quintessence matter  \cite{Sareh, Hajebrahimi}} Recently, Haldar et al. revealed that the geometric and thermodynamic volumes of the Bardeen anti de Sitter black hole surrounded by the quintessence matter are not equal to each other \cite{Haldar2020}.
Ndogmo \emph{et al.}
examined phase transition
in a rotating non-linear magnetic-charged black hole surrounded by the quintessence matter via evaluating the thermodynamic functions \cite{Ndogmo2021}. Last year, we investigated the quintessence matter surrounded Schwarzschild black hole thermodynamics in a deformed Heisenberg algebra where a minimal length exists \cite{BCL1}. Inspired from this article, Hao et al. handled a similar study for the Reissner- Nördstrom black hole within another deformed algebra scenario \cite{BCL2}. 

As we explained above, modified dispersion relations and quintessence matter which surrounds black holes are thought to have important influences on the thermodynamics of black holes. In literature, to our best knowledge, both phenomena impacts on black hole thermodynamics have not been explored simultaneously. With this motivation, we decided to investigate thermal quantities of Schwarzschild black hole surrounded by quintessence matter in the Planck scale by considering a particular form of MDR. We build the manuscript as follows: In the next section we present a brief review of the considered MDR and Schwarzschild black hole. Then, in section 3, we derive the thermodynamics functions of the black hole in the presence of quintessence matter and MDR. The manuscripts ends with the conclusion section.

\section{A brief review of the MDR and Kiselev's black hole} \label{sec2}

In this manuscript, we consider a MDR of the type \begin{equation}
p^{2}=E^{2}+\eta E^{4}-\mu ^{2},  \label{type}
\end{equation}
in natural units,  $\hbar=c=1$, \cite{Camelia}. Here, $\mu$ indicates the mass parameter which is proportional to the rest energy, where $\mu \neq m$. The parameter $\eta$ is defined by $\eta =\beta \ell _{p}^{2}$,   while $\beta $ takes different values in distinct quantum-gravity models \cite{Andrea,Nasser}. A modification in
the dispersion relation implies an alteration between the particle's position and energy uncertainty. To obtain this modification, we employ a variation to the Eq. \eqref{type} and keep only the terms up to the second power of the Planck length 
\begin{equation}
\delta p\simeq \left( 1+\frac{3}{2}\eta E^2\right) \delta E,
\end{equation}%
while we set $\mu =0$.  By assuming $E\simeq \delta E$, we apply the conventional Heisenberg's
uncertainty principle, $\delta E\geq \frac{1}{\delta x}$ and $\delta p\geq 
\frac{1}{\delta x}$; and {\color{red}we obtain modified} the uncertainty relation in the form of
\begin{equation}
E \geq \frac{1}{\delta x}-\frac{3\eta }{2 \left( \delta x\right) ^3}.  \label{B}
\end{equation}%
On the other hand, in 2003 Kiselev examined static spherically symmetric black hole geometry surrounded by quintessence matter using the exact solution of Einstein's field equations. %Kiselev examined the geometry of the static spherically symmetric black hole surrounded by the quintessence matter from the exact solution of Einstein's field equations. 
He showed that the metric 
\cite{Kiselev}
\begin{equation}
ds^{2}=-g\left( r\right) dt^{2}+g^{-1}\left( r\right) dr^{2}+r^{2}d\Omega .
\end{equation}%
corresponds to the Schwarzschild black hole if the lapse function, $g\left( r\right) 
$, is taken as%
\begin{equation}
g\left( r\right) =1-\frac{2M}{r}-\frac{\alpha }{r^{3\omega _{q}+1}}.
\end{equation}%
Here, $M$ represents the black hole mass, $\alpha $ denotes a positive normalization
factor depending on  quintessence matter, and $\omega _{q}$ indicates the quintessential state
parameter which receives values in the range  of {\color{red}$-1< \omega _{q}\leq -1/3$}. The roots of lapse function define the outer and inner horizons
\begin{equation}
\left. 1-\frac{2M}{r}-\frac{\alpha }{r^{3\omega _{q}+1}}\right \vert
_{r=r_{H}}=0. \label{Mm}
\end{equation}
We notice that the quintessence state parameter determines the degree of the root, therefore, based on some particular values of the quintessential state parameter, we can categorize the solutions to the following three cases:
\begin{itemize}
\item The $\omega _{q}=-2/3$ case,  black hole gets two horizons, namely inner and outer horizons, in the form of%
\begin{equation}
r_{in}=\frac{1-\sqrt{1-8\alpha M}}{2\alpha },\quad\quad r_{out}=\frac{%
1+\sqrt{1-8\alpha M}}{2\alpha },\quad \text{ with } \quad 8\alpha M < 1.
\end{equation}

\item The $\omega _{q}=-1/3$ case, black hole has only one horizon,%
\begin{equation}
r_{in}=\frac{2M}{1-\alpha }.
\end{equation}

\item The $\omega _{q}=-1$ case, the metric reduces to the Schwarzschild de-Sitter
metric, if%
\begin{equation}
\alpha \simeq \frac{\Lambda }{3}.
\end{equation}
\end{itemize}

\section{Black hole thermodynamics with the MDR} \label{sec3}

In this section, we investigate the thermodynamics of the Schwarzschild black hole  surrounded by quintessence matter in the presence of the given MDR in the previous section. To this end first, we calculate the entropy function by following the heuristic consideration of Bekenstein \cite{Bek}. According to his approach, when the black hole absorbs a classical particle with energy, $E$, and size, $R$, its area increases minimally as
\begin{equation}
\left( \Delta A\right) _{\min }\geq 4\left( \ln 2\right) \ell _{p}^{2}ER.
\end{equation}
However, at the quantum mechanical level the size of the particle can be described in terms of the uncertainty in its position $R\simeq \delta x\simeq r_{H}$, and therefore, the minimal increase can be given as
\begin{equation}
\left( \Delta A\right) _{\min }\geq 4\left( \ln 2\right) \ell
_{p}^{2}E\delta x.
\end{equation}%
Accommodating the Planck-length modification,  given in Eq. \eqref{B}, we obtain%
\begin{equation}
\left( \Delta A\right) _{\min }\geq 4\ell _{p}^{2}\left( \ln 2\right) \left(
1-\frac{3\eta }{2\left( \delta x\right) ^{2}}\right) .
\end{equation}%
Setting a minimal increase in entropy, namely by taking $\left( \Delta S\right) _{\min }=\ln 2$, we write%
\begin{equation}
\frac{dS}{dA}\simeq \frac{\left( \Delta S\right) _{\min }}{\left( \Delta
A\right) _{\min }}\simeq \frac{1}{4\ell _{p}^{2}}\left( 1+\frac{3\eta }{%
2r_{H}^{2}}\right), \label{C}
\end{equation}
which leads entropy to take the form of:
\begin{equation}
S=\frac{A}{4\ell _{p}^{2}}+\frac{3\pi \eta }{2\ell _{p}^{2}}\log A.
\end{equation}%
We observe that the quintessence matter does not alter the black hole's entropy, while
the MDR correction does. It is of interest to note that the {\color{red}logarithmic correction term} stands in the entropy function as it appears in many
theories of quantum gravity {\color{red}\cite{BCL2022}. However, depending on the considered approaches this correction term can be positive or negative. In a featured article, Nozari et al. showed that the correction term takes a positive sign in MDR approach due to the absence of odd power energy terms \cite{App}. They revealed that this sign is negative in another approach, namely in the generalized uncertainty principle (GUP) approach, where even power of $\Delta x$ do not exist. In \cite{Sabina06}, Hossenfelder noted a close-relation between MDR and GUP approaches. In the current problem, if we can set $\frac{3\eta }{2\ell _{p}^{2}}=-\beta$ to demonstrate this correlation. Note that here $\beta$ denotes the deformation parameter of the GUP, and  for $\eta=0$, the logarithmic correction term vanishes, thus,} we obtain the conventional form of the entropy function.

Next, we intend to calculate the corrected Hawking temperature. For that, we employ the first law of black-hole thermodynamics \cite{Haw1, Haw2}
\begin{equation}
dS=\frac{dM}{T}.
\end{equation}
Following the straightforward algebra, we obtain the MDR-corrected Hawking temperature of black hole in terms of horizon radius as follows:%
\begin{equation}
T_{H}=\frac{\ell _{p}^{2}}{4\pi r_{H}}\left( 1+\frac{3\alpha \omega _{q}}{%
r_{H}^{3\omega _{q}+1}}\right) \left( 1-\frac{3\eta }{2r_{H}^{2}}\right) .
\label{tem}
\end{equation}%
For $\eta =0$, Eq. \eqref{tem} reduces to the given form 
\begin{equation}
T_{H}=\frac{\ell _{p}^{2}}{4\pi r_{H}}\left( 1+\frac{3\alpha \omega _{q}}{%
r_{H}^{3\omega _{q}+1}}\right), \label{Theta}
\end{equation}%
as found in \cite{BCL1}. Furthermore, in the limit of $\alpha \rightarrow 0$, the second term on Eq. \eqref{Theta} drops and we arrive at the conventional Hawking temperature of the
Schwarzschild black hole. Next, we analyze the MDR-corrected Hawking temperature at
particular quintessence state parameter values:

\begin{itemize}
\item In the $\omega _{q}=-2/3$ case, the MDR-corrected Hawking temperature becomes
\begin{equation}
T_{H}=\frac{\ell _{p}^{2}}{4\pi r_{H}}\left( 1-2\alpha r_{H}\right) \left( 1-%
\frac{3\eta }{2r_{H}^{2}}\right) .
\end{equation}%
To get a physical temperature, we must impose $\left( 1-2\alpha r_{H}\right) \left( 1-\frac{3\eta }{2r_{H}^{2}}\right) \geq 0$. This requirement can be met if both multipliers are positive or negative at the same time. We note that these conditions specify different bound values on horizon radius. For example, if both multipliers are positive, then  we get
\begin{equation}
\frac{1}{2\alpha }\geq r_{H}\geq \sqrt{\frac{3\eta }{2}}. \label{bound1}
\end{equation}
while if both of them are negative, we find 
\begin{equation}
\sqrt{\frac{3\eta }{2}} \geq r_{H}\geq \frac{1}{2\alpha }.
\end{equation}
However, the later solution must be omitted because, in the $\eta=0$ limit, it is unphysical.  It is worth mentioning that in the physical solution,  the Hawking temperature goes to zero at two horizon values, here $\frac{1}{2\alpha }$ and {\color{red}$\sqrt{\frac{3\eta}{2}}$}, remnants can be expected. We will discuss this in more detail later when we examine the heat capacity function.

\end{itemize}
%and $4M < \frac{1}{2\alpha}$ should be satisfied.

\begin{itemize}
\item In the $\omega _{q}=-1$ case, the MDR-corrected Hawking temperature
takes up the form 
\begin{equation}
T_{H}=\frac{\ell _{p}^{2}}{4\pi r_H}\left( 1-3\alpha r_{H}^{2}\right) \left( 1-%
\frac{3\eta }{2r_{H}^{2}}\right),
\end{equation}%
while the horizons radius can vary in the range%
\begin{equation}
\frac{1}{\sqrt{3\alpha }}\geq r_{H}\geq \sqrt{\frac{3\eta }{2}}. \label{bound2}
\end{equation}

\item In the $\omega _{q}=-1/3$ case, Eq. (\ref{tem}) stands as%
\begin{equation}
T_{H}=\frac{\ell _{p}^{2}}{4\pi r_{H}}\left( 1-\alpha \right) \left( 1-\frac{%
3\eta }{2r_{H}^{2}}\right) ,
\end{equation}
and the horizon radius is bounded only from below,%
\begin{equation}
r_{H}\geq \sqrt{\frac{3\eta }{2}}. \label{bound3}
\end{equation}%
\end{itemize}
Therefore, we conclude that the MDR-correction states the same lower bound on the horizon, while the upper bound of the horizon is specified by the quintessence matter state parameter. 

\begin{figure*}[tbh]
\resizebox{\linewidth}{!}{\includegraphics{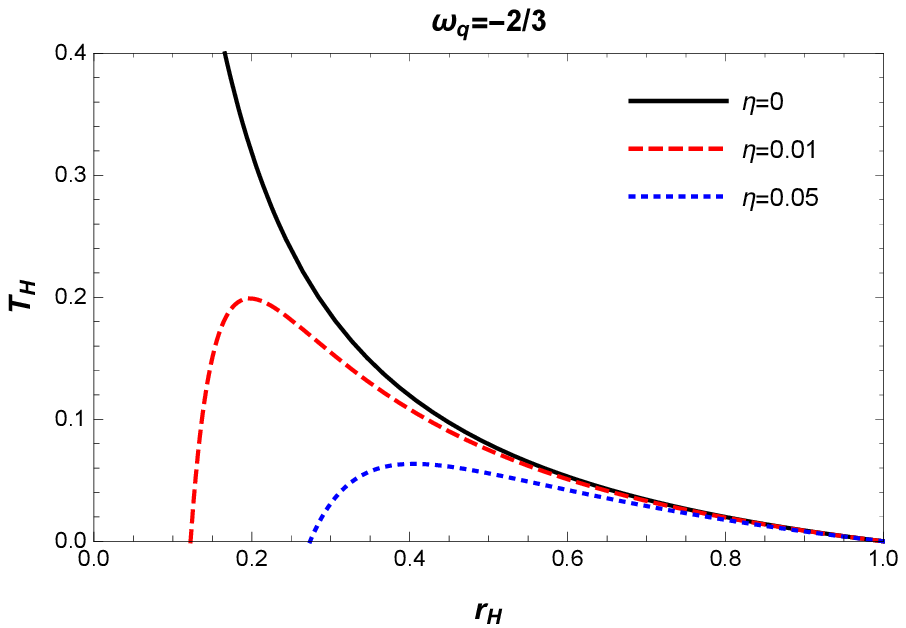},\includegraphics{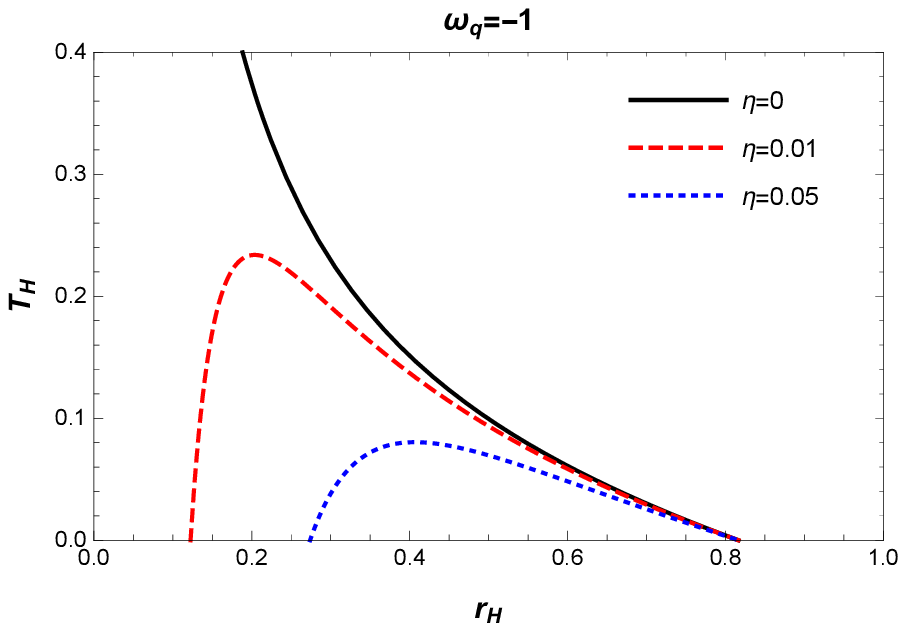},%
\includegraphics{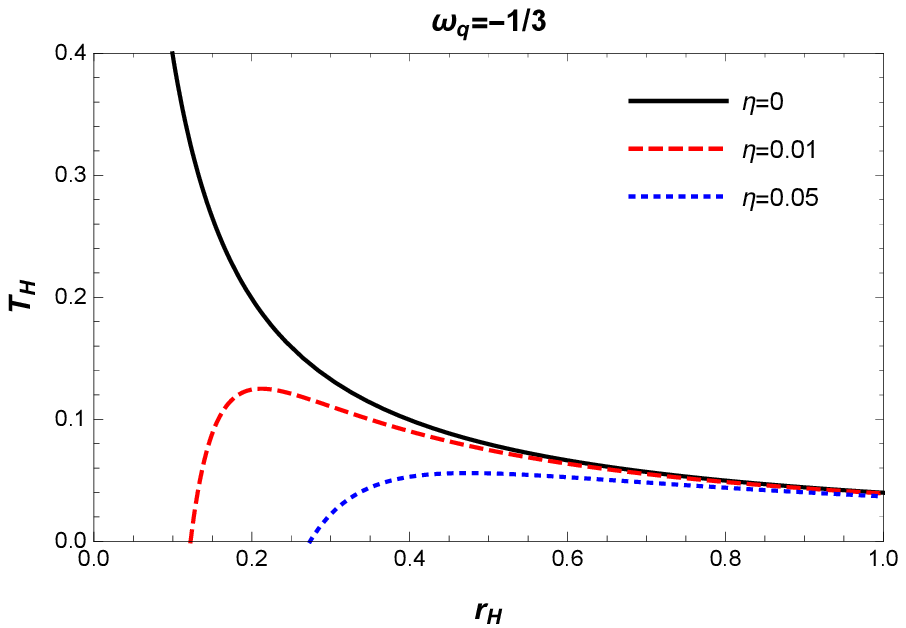}}
\caption{MDR-corrected Hawking temperature versus horizon radius.}
\label{Fig1}
\end{figure*}

In the rest of manuscript, we demonstrate our findings for $\omega _{q}=-2/3$, $\omega _{q}=-1$, and $\omega _{q}=-1/3$ cases by taking  $\eta =0$, $\eta =0.01$, $\eta =0.05$, that correspond to unmodified and modified dispersion relation, for $\ell _{p}=1,$  $\alpha =0.5$. First, in Fig. \ref{Fig1}, we present the behavior of the MDR-corrected Hawking temperature versus the horizon radius.

In all cases of $\eta =0$, we observe that  the Hawking temperature decreases smoothly while the horizon radius increases. This behavior is not valid for the $\eta \neq 0$ cases. The MDR-corrected Hawking temperature grows very fast with increasing horizon radius and after gaining a maximum or a saturated value it varies smoothly with increasing $r_{H}$.
We see that for $\eta =0$, $\eta =0.01$, $\eta =0.05$, horizons have the same lower bound values: $0$, $0.122$, and $0.274$, respectively. The upper value of horizons are 
equal to $1$, $0.816$ and infinity for
the $\omega _{q}=-2/3$, 
$\omega _{q}=-1/3$ and 
$\omega _{q}=-1$ cases, respectively, as predicted by the Eq. \eqref{bound1}, Eq. \eqref{bound2}, and Eq. \eqref{bound3}.

\begin{figure*}[tbh]
\resizebox{\linewidth}{!}{\includegraphics{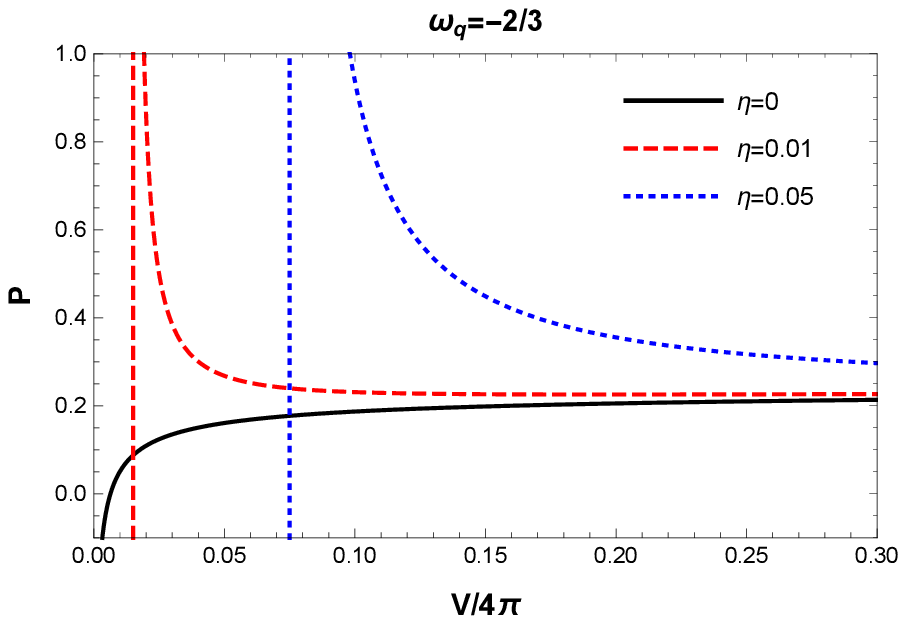},\includegraphics{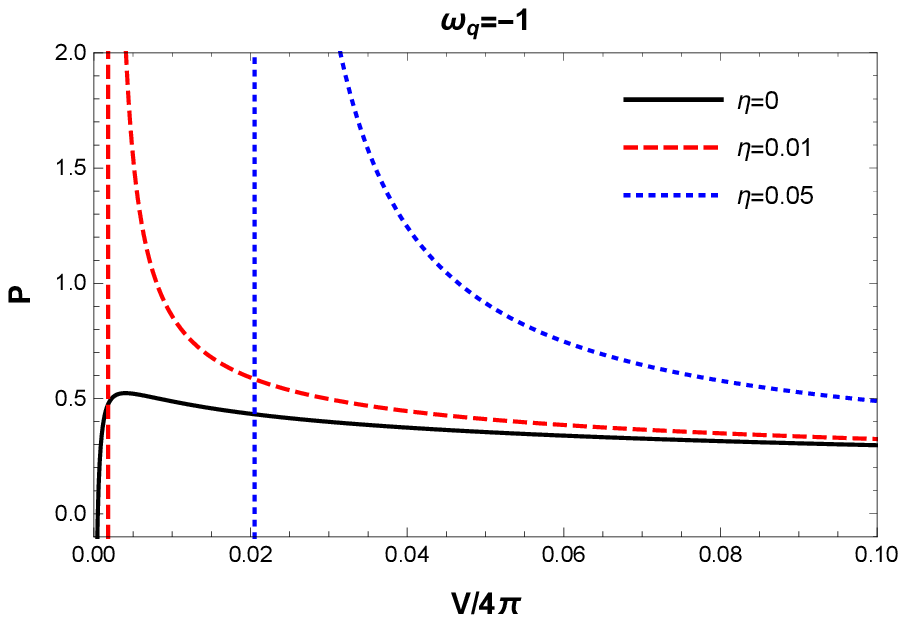},%
\includegraphics{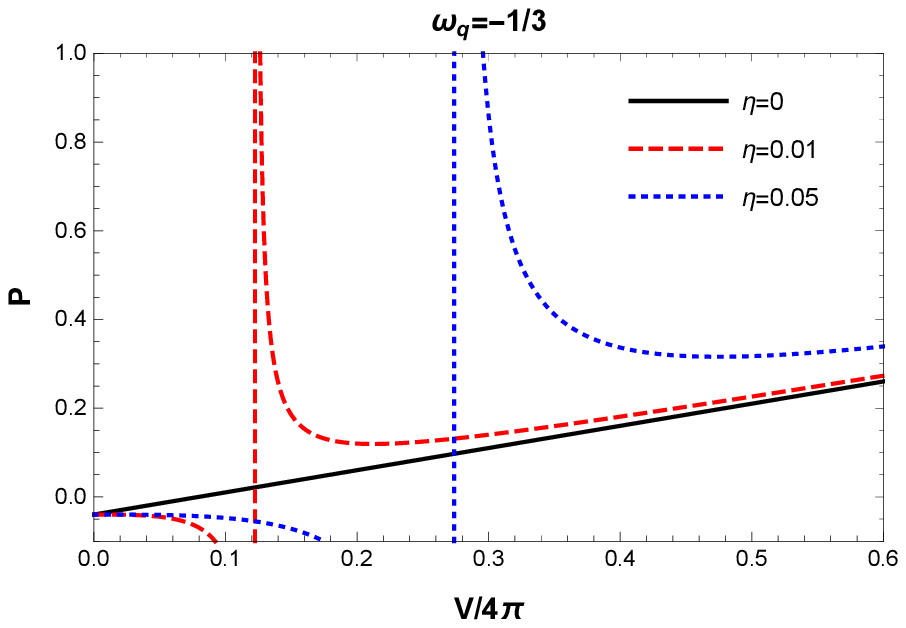}} 
\caption{MDR corrected P-V isotherms for $T_H=1$.}
\label{P3}
\end{figure*}

{\color{red}The quintessence matter can be thought of a perfect fluid \cite{Kastor}.}
Next, we deduce black hole's  equation of state function by using the relation between the pressure $P$ and $\alpha$ parameter {\color{red}out of stress-energy tensor,}
\begin{equation}
P=-\frac{\alpha }{8\pi }. \label{pres}
\end{equation}
We substitute it in Eq. \eqref{Mm} to associate the mass with the enthalpy. We find
\begin{equation}
M=H=\frac{r_{H}}{2}+%
\frac{4\pi }{r_{H}^{3\omega _{q}}}P.
\end{equation}%
Then, by using Legendre transformation, we obtain the volume of the black hole as%
\begin{equation}
V=\left( \frac{\partial M}{\partial P}\right) _{S}=\frac{4\pi }{%
r_{H}^{3\omega _{q}}}.
\end{equation}
By using this relation, we express the horizon radius in terms of the volume
\begin{equation}
r_{H}=\left( \frac{4\pi }{V}\right) ^{\frac{%
1}{3\omega _{q}}}. \label{rvol}
\end{equation}%
Since horizons have upper and lower bound values, we can specify the same for the volume. For the particular quintessence state parameters, the volume and their changeable physical range stand as follows:
\begin{itemize}
    \item For $\omega _{q}=-2/3$:
    \begin{eqnarray}
    V=4\pi r_H^2,
    \end{eqnarray}
and
\begin{subequations} \label{Vbor1}
\begin{eqnarray} 
&& 1 \geq \frac{V}{4\pi}\geq 0, \,\,\,\, \quad \quad \quad \eta=0,\\
&& 1 \geq \frac{V}{4\pi}\geq 0.015, \quad \quad \eta=0.01,\\
&& 1 \geq \frac{V}{4\pi}\geq 0.075, \quad \quad \eta=0.05.
\end{eqnarray}
\end{subequations}

    \item For $\omega _{q}=-1$:
    \begin{eqnarray}
    V=4\pi r_H^3,
    \end{eqnarray}
and
\begin{subequations} \label{Vbor2}
\begin{eqnarray} 
&& 0.544 \geq \frac{V}{4\pi}\geq 0,  \,\,\,\, \quad \quad \quad \eta=0,\\
&& 0.544 \geq \frac{V}{4\pi}\geq 0.002, \quad \quad \eta=0.01,\\
&& 0.544 \geq \frac{V}{4\pi}\geq 0.021, \quad \quad \eta=0.05.
\end{eqnarray}
\end{subequations}
    
    \item For $\omega _{q}=-1/3$:
    \begin{eqnarray}
    V=4\pi r_H,
    \end{eqnarray}
and
\begin{subequations} \label{Vbor3}
\begin{eqnarray} 
&& \frac{V}{4\pi}\geq 0, \,\,\,\,  \quad \quad \quad \eta=0,\\
&&  \frac{V}{4\pi}\geq 0.122, \quad \quad \eta=0.01,\\
&&  \frac{V}{4\pi}\geq 0.274, \quad \quad \eta=0.05.
\end{eqnarray}
\end{subequations}    
\end{itemize}
Then, by using Eqs. \eqref{pres} and \eqref{rvol} in Eq. \eqref{tem}, we express the black hole's equation of state function as follows:
\begin{equation}
P=\frac{1}{24\pi \omega _{q}}\left( \frac{4\pi }{V}\right) ^{1+\frac{1}{3\omega _{q}}}\left \{ 1-\frac{4\pi }{\ell _{p}^{2}}\left( \frac{%
4\pi }{V}\right) ^{\frac{1}{3\omega _{q}}}\frac{T_{H}}{\left( 1-\frac{3\eta 
}{2}\left( \frac{4\pi }{V}\right) ^{-\frac{2}{3\omega _{q}}}\right) }\right
\} .
\end{equation}%
For the particular quintessence state parameters, equation of state stands as follows:
\begin{itemize}
\item $\omega _{q}=-2/3$:
\begin{equation}
P=-\frac{1}{16\pi}\left( \frac{4\pi }{V}\right) ^\frac{1}{2}\left \{ 1-\frac{4\pi }{\ell _{p}^{2}}\left( \frac{%
4\pi }{V}\right) ^{-\frac{1}{2}}\frac{T_{H}}{\bigg( 1-\frac{3\eta }{2}\left( \frac{4\pi }{V}\right) \bigg) }\right
\} . \label{PV1}
\end{equation}%

\item $\omega _{q}=-1$:
\begin{equation}
P=-\frac{1}{24\pi}\left( \frac{4\pi }{V}\right) ^{\frac{2}{3}}\left \{ 1-\frac{4\pi }{\ell _{p}^{2}}\left( \frac{%
4\pi }{V}\right) ^{-\frac{1}{3}}\frac{T_{H}}{\bigg( 1-\frac{3\eta }{2}\left( \frac{4\pi }{V}\right) ^{\frac{2}{3}}\bigg) }\right
\} . \label{PV2}
\end{equation}%

\item $\omega _{q}=-1/3$:
\begin{equation}
P=-\frac{1}{8\pi}\left \{ 1-\frac{4\pi }{\ell _{p}^{2}}\left( \frac{%
4\pi }{V}\right) ^{-1}\frac{T_{H}}{\bigg( 1-\frac{3\eta 
}{2}\left( \frac{4\pi }{V}\right) ^{2}\bigg) }\right
\} . \label{PV3}
\end{equation}%
\end{itemize}
In order to present the impact of the deformation parameters on the $P-V$
isotherm, we depict Eq. \eqref{PV1}, Eq. \eqref{PV2}, and Eq. \eqref{PV3} versus reduced volume in Fig. \ref{P3}. 

We observe a singularity in the lower physical value of the reduced volume, which confirm  Eq. \eqref{Vbor1},
Eq. \eqref{Vbor2}, and Eq. \eqref{Vbor3} for  
$\omega_q=-2/3$, $\omega_q=-1$, and $\omega_q=-1/3$ cases, respectively. The MDR-corrected P-V isotherms differ from the unmodified ones, especially in the smaller reduced volume. This difference becomes greater for higher values of $\eta$ parameter.

Finally, we deduce the modifed heat capacity function to discuss the thermal stability of the considered black hole. We employ
\begin{equation}
C=\frac{dM}{dT},
\end{equation}%
and obtain the following general expression%
\begin{equation}
C=-\frac{2\pi r_{H}^{2}}{\ell _{p}^{2}}\frac{\left( 1+\frac{3\alpha \omega
_{q}}{r_{H}^{3\omega _{q}+1}}\right) }{1+\frac{3\alpha \omega _{q}\left(
3\omega _{q}+2\right) }{r_{H}^{3\omega _{q}+1}}-\frac{9\eta }{2r_{H}^{2}}\left( 1+\frac{\alpha
\omega _{q}\left(3 \omega _{q}+4\right) }{r_{H}^{3\omega _{q}+1}}\right) 
}.  \label{heat}
\end{equation}
Then, we express the MDR-corrected heat capacity for some particular
quintessence state parameter as we have done above. 
\begin{itemize}
\item For $\omega _{q}=-2/3:$ Eq. (\ref{heat}) reduces to
\begin{equation}
C=-\frac{2\pi r_{H}^{2}}{\ell _{p}^{2}}\frac{\left( 1-2\alpha r_{H}\right) }{%
1-\frac{9\eta }{2r_{H}^{2}}\left(1-\frac{4\alpha}{3} r_{H}\right) }.  \label{heat1}
\end{equation}
In this case, the MDR-corrected heat capacity function goes to zero at $r_{H}=1/(2\alpha) $ which indicates 
the existence of a remnant 
\begin{eqnarray}
M_{\text{rem}}=\frac{1}{8\alpha},
\end{eqnarray}
since the black hole cannot exchange radiation with its surrounding. In the physical interval, we observe a singularity at 
\begin{eqnarray}
r_{H_{CR}}=3\eta \alpha \left(\sqrt{1+\frac{1}{4\eta \alpha^2}}-2\right).
\end{eqnarray}
For $\eta =0$, $\eta =0.01$, $\eta =0.05$, this critical horizon radius is equal to $0$, $0.198$ and $0.405$ for $\alpha=0.5$.

\item For $\omega _{q}=-1:$ Eq. (\ref{heat}) simplifies to%
\begin{equation}
C=-\frac{2\pi r_{H}^{2}}{\ell _{p}^{2}}\frac{\left( 1-3\alpha
r_{H}^{2}\right) }{ 1+3\alpha r_{H}^{2}-\frac{9\eta}{2r_H^2} \left( 1-\alpha
r_{H}^{2}\right) }.
\end{equation}%
In this case, the heat capacity goes to zero at $r_{H}=1/\sqrt{3\alpha }$ confirming the presence of a remnant. 
\begin{eqnarray}
M_{\text{rem}}=\frac{1}{3\sqrt{3\alpha }}.
\end{eqnarray}
We find a critical horizon value at
\begin{eqnarray}
r_{H_{CR}}=\sqrt{\frac{-(2+9\eta \alpha)+ \sqrt{(2+9\eta \alpha)^2+216 \eta \alpha}
}{12 \alpha}},
\end{eqnarray}
which gives $0$, $0.204$ and $0.407$ for $\eta =0$, $\eta =0.01$, $\eta =0.05$, respectively when $\alpha=0.5$.

\item $\omega _{q}=-1/3:$ Eq. (\ref{heat}) gives%
\begin{equation}
C=-\frac{2\pi r_{H}^{2}}{\ell _{p}^{2}}\frac{1}{1-\frac{\eta }{2r_{H}^{2}}},
\end{equation}%
that is $\alpha$  independent. It is worth mentioning that in this case there is no remnant in the physical region. We observe singularity at 
\begin{eqnarray}
r_{H_{CR}}=\sqrt{\frac{9 \eta}{2} },
\end{eqnarray}
which is equal to  $0$, $0.212$, and $0.474$  for $\eta =0$, $\eta =0.01$, $\eta =0.05$, respectively.
\end{itemize}

\begin{figure*}[tbh]
\resizebox{\linewidth}{!}{\includegraphics{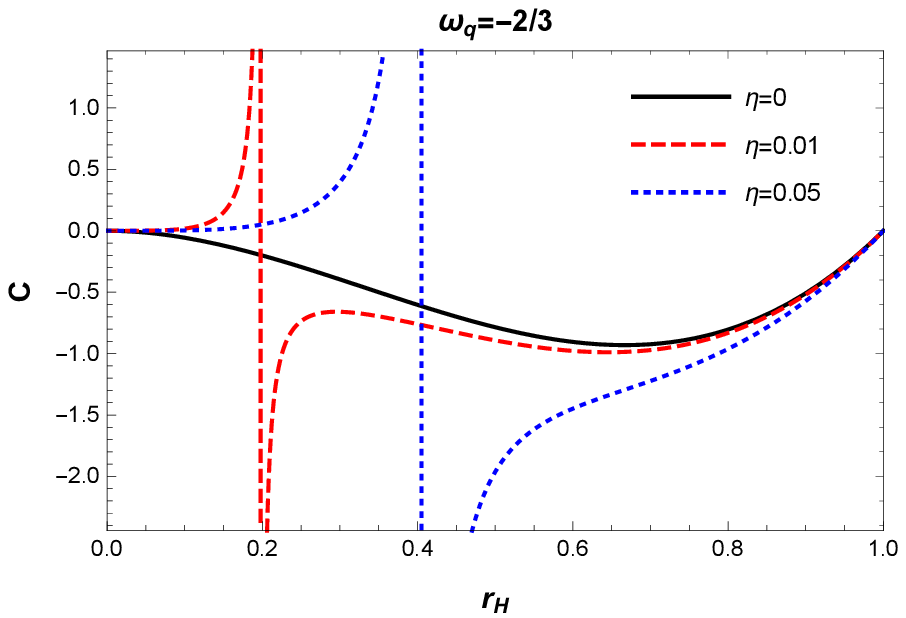},\includegraphics{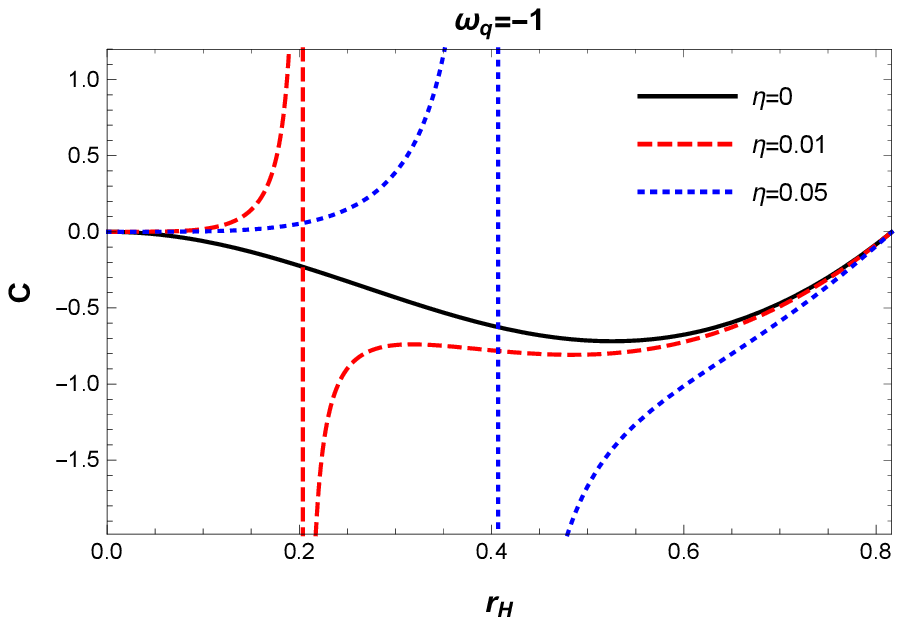},%
\includegraphics{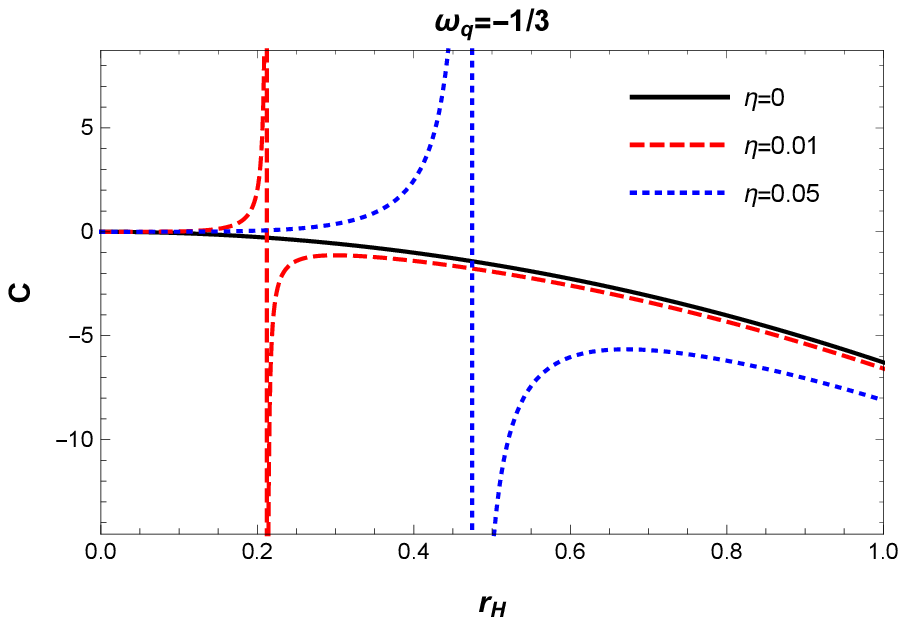}}
\caption{MDR-corrected specific heat versus horizon radius.} \label{SH}
\end{figure*}

For three particular cases, we illustrate the plots of the MDR-corrected heat capacity function versus the horizon radius in Fig. \ref{SH}. 

We observe that the MDR plays an important role in the stability of black hole. When $\eta=0$, the black holes are unstable in all cases. However, when $\eta \neq 0$, black hole becomes stable for $r_H < r_{H_{CR}}$ and unstable for  $r_H > r_{H_{CR}}$. Therefore, $r_H = r_{H_{CR}}$ points represent the phase transition point of the black hole. We note that these discontinuity points are the turning points of the MDR-corrected Hawking radiation given in Fig. \ref{Fig1}.

\section{Conclusion}
In this manuscript, we investigate the thermodynamics of Schwarzschild's black hole embedded in the quintessence matter in the context of a modified dispersion relation. At first, with the heuristic approach, we obtain the modified entropy. We show that the logarithmic area term arises with the MDR correction, and the total entropy is not affected by the quintessence matter. Then, we examine the MDR-corrected Hawking temperature. We find that the physically acceptable Hawking temperature concept
requires additional conditions on the horizon. We show that a lower and an upper bound on the horizon radius stem from the modified dispersion and the quintessence matter, respectively.
Then, we express the volume in terms of the horizon. After that, we derive the MDR-corrected equation of state function. We observe the effect of the MDR-corrections on the  P-V isotherms at lower values of the reduced volume. Finally, we deduce the specific heat function. We see that a remnant can occur depending on the Lovelock gravity parameter. Moreover, we find that there is a critical horizon value, which depends on the MDR parameter. These points are the turning points of the Hawking temperature, and they represent a phase transition that makes the black hole become unstable.

\section*{Acknowledgments}
{\color{red}The authors thank the anonymous referee for a thorough reading of our
manuscript and for constructive suggestions.} BCL is supported by the Internal  Project,  [2022/2218],  of  Excellent  Research  of  the  Faculty  of  Science  of Hradec Kr\'alov\'e University.

\section*{Data Availability Statements}
The authors declare that the data supporting the findings of this study are available within the article.

\end{document}